\documentclass[aip, preprint,showpacs,showkeys]{revtex4-1}
\usepackage{amssymb, dcolumn,  graphicx, latexsym, amsfonts, bm}

\begin{document}

\title{Non-modal kinetic theory of the hydrodynamic drift instabilities 
of  plasma shear flows}
\author{V. V. Mikhailenko}
\affiliation{Department of Electrical Engineering, Pusan National University, Busan 609--735, S. Korea.}
\email[E-mail: ]{vladimir@pusan.ac.kr}
\author{V. S. Mikhailenko}
\affiliation{Department of Physics and Technology, V.N. Karazin Kharkov National University, 
61108 Kharkov, Ukraine.}
\author{Hae June Lee}
\affiliation{Department of Electrical Engineering, Pusan National University, Busan 609--735, S. Korea.}

\date{\today}

\begin{abstract}
The non-modal kinetic theory of the kinetic drift instability of plasma shear flows 
[Phys.Plasmas, 18, 062103 (2011)] is extended to the investigation of the long-time 
evolution of the hydrodynamic ion temperature gradient and resistive drift instabilities 
in plasma shear flow. We find, that these hydrodynamic instabilities passed in their
temporal evolution in shear flow through the kinetic stage of the evolution. In linear 
theory, this evolution involves the time dependent, due to flow shear, effects of the 
finite Larmor radius, which resulted in the non-modal effect of the decrease with 
time the frequencies and the growth rates of the instabilities.
\end{abstract}

%\keywords{diocotron instability; non-neutral plasma}
\maketitle

%\preprint{APS/123-QED}

\altaffiliation{Research Institute of Computer, Information and Communication, 
Pusan National University, 30 Jangjeon-dong, Guemjeong-gu, Pusan 609--735, S. Korea.} 

\affiliation{Department of Physics and Technology, V.N. Karazin Kharkov National University, 
61108 Kharkov, Ukraine.}

\section{INTRODUCTION}

The investigation of the long-time mode of behaviour of the regimes
of tokamak plasma improved confinement, where E x B flow
shear stabilization mechanisms \cite{burell} plays a key role, is
the theme of major importance in the physics of continuous tokamaks
operation. Gyrokinetic theory \cite{Brizard-Hahm} gives a foundation
for the investigating microinstabilities, which cause the anomalous
transport in fusion plasmas. While there has been significant
progress in gyrokinetic treatment of tokamak turbulence (see, for
example, Refs.\cite {Brizard}-\cite {Rewoldt}), there key issues still remain
in application of the gyrokinetic theory to analytical investigations of the 
long-time evolution of plasma shear flows turbulence.

In Ref.\cite{Mikhailenko-2011} we have developed new kinetic description of the plasma shear flows, 
using the Kelvin's method of shearing modes or so-called non-modal approach as its foundation. 
The governing equation in that theory is the 
integral equation for the perturbed electrostatic potential, which is the formal solution of 
the initial value problem for Vlasov-Poisson (V-P) system.  
The important procedure in its derivation is the transformation of  V-P system 
to sheared (in space and velocity) coordinates convected with flow and
accounting for by this mean the effect of spatial time-dependent distortion 
of plasma disturbances by shear flows. In that equation, velocity 
shear reveals as the non-modal time-dependent effect of the finite Larmor radius.
Because of the shearing of perturbations in shear flow, 
the component of the wave number along the direction of the velocity shear experiences 
the growth with time, and therefore results obtained for the hydrodynamic drift 
instabilities on the base of the fluid description are valid only for finite times 
till $k_{\perp}\rho_{i}\ll 1$ is valid ( $\rho_{i}$ is thermal ion Larmor radius). 
Therefore, not only kinetic, but also typically hydrodynamic instabilities 
of shearless plasma require the kinetic description for plasma shear flows for proper treating their 
evolution on times, at which $k_{\perp}\rho_{i}$  becomes not small and approaches unity and above. 
It reveals\cite{Mikhailenko-2011}, that conventional kinetic theory, as well as the gyrokinetic theory, 
in which the time dependence of the 
perturbed distribution functions and fields are considered in a canonical modal 
form, $\varphi\backsim \exp\left(-i\omega t\right)$, are valid only at the 
initial stage of the shear flow turbulence evolution, 
at times less then the inverse velocity shearing rate $\left(V'_{0}\right)^{-1}$. 
The theory developed displays that the experimentally observed suppression 
of drift-type turbulence and the improved energy confinement 
are appreciable at $t\gtrsim \left(V'_{0}\right)^{-1}\gtrsim 
\left(\gamma\right)^{-1}$, i.e. just after that time.  In our analysis of the kinetic 
drift instability we have obtained \cite{Mikhailenko-2011}, that in linear theory of the kinetic 
(universal) drift instability, shear flow leads at time $t\gtrsim \left(V'_{0}\right)^{-1}$ to the 
non-modal decrease with time the frequency and growth rate and to ultimate suppression of that instability.

The basic transformation of the V-P system to sheared coordinates, which resulted 
in the removing from the Vlasov equation the term, which contains the spatial inhomogeneity introduced 
by the velocity shear, is presented and discussed in Sec. II.  
In Sec. III, we apply the linear non-modal kinetic theory of 
Ref.\cite{Mikhailenko-2011}, extended onto the accounting for the inhomogeneity 
of the ion temperature, to the investigation of the temporal evolution of the 
hydrodynamic ion temperature gradient drift instability 
of plasma shear flow. We find that the ordinary modal theory of this instability is valid only for 
times less than the inverse velocity shearing rate. At times $t_{s} 
\gg t\gtrsim \left(V'_{0}\right)^{-1}$, where $t_{s}=\left(V'_{0}k_{y}\rho_{i}\right)^{-1}$, 
the non-modal effects of the decrease of the frequency and growth rate develop. 
At the final stage, at times $t\gg t_{s}$ the perturbations of the cell type with zero frequency occurs.

In Sec.IV, using the integral equation for electrostatic potential obtained in Ref.\cite{Mikhailenko-2011}, 
extended onto the accounting for the collision of electrons with neutrals, we consider temporal 
evolution of the resistive drift instability of plasma with comparable ion and electron temperatures, 
as it is at the edge layer of tokamaks. We  came to the same conclusion about the importance of the 
kinetic theory for the proper treating of the resistive drift instability in shear flow. It is shown, 
that the reducing with time the frequency and growth rate is a common future of the 
evolution of the electrostatic drift instabilities in plasma shear flows. 
A summary of the work is given in Conclusions, Section V.

\section{VLASOV--POISSON SYSTEM OF EQUATIONS \newline IN SHEARED COORDINATES}
We start with the Vlasov equation for species $\alpha$ ($\alpha=i$ for ions 
and $\alpha=e$ for electrons), 
immersed in crossed  spatially inhomogeneous electric field,
$\mathbf{E}_{0}\left( \hat{\mathbf{r}}\right)$
and homogeneous magnetic field $\mathbf{B}\parallel\mathbf{e}_{z}$,
\begin{eqnarray}
&\displaystyle \frac{\partial F_{\alpha}}{\partial
t}+\hat{\mathbf{v}}\frac{\partial F_{\alpha}}
{\partial\hat{\mathbf{r}}}+\frac{e}{m_{\alpha}}\left(\mathbf{E}_{0}
\left(\hat{\mathbf{r}}\right)
+\frac{1}{c}\left[\hat{\mathbf{v}}\times\mathbf{B}\right]
-\bigtriangledown
\varphi\left(\hat{\mathbf{r}},t\right)\right)\frac{\partial
F_{\alpha}}{\partial\hat{\mathbf{v}}}=0.\label{1}
\end{eqnarray}
We use a slab geometry with the mapping $\left(r,\theta,
\varphi\right)\rightarrow \left(\hat{x}, \hat{y}, \hat{z}\right)$ where $r,\theta,
\varphi$ are the radial, poloidal and toroidal directions,
respectively, of the toroidal coordinate system. In this paper, we consider the 
case of plasma shear flow in linearly changing electric field, $\mathbf{E}_{0}
\left( \hat{\mathbf{r}}\right)= \left( \partial
E_{0}/\partial \hat{x}\right) \hat{x}\mathbf{e}_{x}$ with  $ \partial
E_{0}/\partial \hat{x}=const$ . In that case
\begin{eqnarray}
&\displaystyle \mathbf{V}_{0}\left(\mathbf{r}
\right)=V_{0}\left(\hat{x}\right)\mathbf{e}_{y} =-\frac{c}{B}\frac{\partial
E_{0}}{\partial \hat{x}}\hat{x}\mathbf{e}_{y} =V'_{0}\hat{x}\mathbf{e}_{y}.\label{2}
\end{eqnarray}
with spatially homogeneous, $V'_{0}=const$, velocity shear. The possible spatially homogeneous part of shear flow velocity is eliminated from the problem by a simple Galilean transformation.
It was obtained in Ref.\cite{Mikhailenko-2011}, that transition in 
the Vlasov equation from velocity $\hat{\mathbf{v}}$  and coordinates $\hat{x}$,~$\hat{y}$,~$\hat{z}$
to convected coordinates $\mathbf{v}$ in velocity space, determined by 
\begin{eqnarray}
&\displaystyle \hat{v}_{x}=v_{x},\,\,  \hat{v}_{y}=v_{y}+V'_{0}x,\,\,  \hat{v}_{z}=v_{z}
\label{3}
\end{eqnarray}
and to sheared with flow coordinates $x$,~$y$,~$z$ in the configurational space, determined by
\begin{eqnarray}
&\displaystyle 
\hat{x}=x, \,\,\,\hat{y}=y+V'_{0}tx,\,\,\hat{z}=z 
\label{4}
\end{eqnarray}
(it is assumed that inhomogeneous electric field, 
and respectively shear flow originate at time $t=t_{\left(0\right)}=0$) transforms the linearized Vlasov 
equation for the perturbation of the distribution function $f_{\alpha}=F_{\alpha}-F_{0\alpha}$, with 
known equilibrium distribution $F_{0\alpha}$, to the form, 
\begin{eqnarray}
&\displaystyle \frac{\partial f_{\alpha}}{\partial t}+v_{\alpha
x}\frac{\partial f_{\alpha}}{\partial x} +\left(v_{\alpha
y}-v_{\alpha x}V'_{\alpha}t \right) \frac{\partial
f_{\alpha}}{\partial y}+ v_{\alpha
z}\frac{\partial f_{\alpha}}{\partial z_{\alpha}} + \omega_{c\alpha} v_{\alpha y}
\frac{\partial f_{\alpha}}{\partial v_{\alpha
x}}-\left(\omega_{c\alpha}+V'_{\alpha} \right) v_{\alpha
x}\frac{\partial f_{\alpha}}{\partial v_{\alpha y}} \nonumber
\\  &\displaystyle
=\frac{e_{\alpha}}{m_{\alpha}}\left(\frac{\partial \varphi}{\partial
x} -V'_{\alpha}t\frac{\partial \varphi}{\partial y} \right)
\frac{\partial F_{0\alpha}}{\partial v_{\alpha x}}
+\frac{e_{\alpha}}{m_{\alpha}} \frac{\partial \varphi}{\partial y}
\frac{\partial F_{0\alpha}}{\partial v_{\alpha y}}
+\frac{e_{\alpha}}{m_{\alpha}} \frac{\partial \varphi}{\partial
z_{\alpha}} \frac{\partial F_{0\alpha}}{\partial v_{\alpha
z}}.\label{5}
\end{eqnarray}
($\omega_{c}$ is the cyclorton frequency of ion (electron)) which is free from the 
spatial inhomogeneities originated from shear flow (see also Eq.(8) in Ref.\cite{Mikhailenko-2011}). 
The Fourier transformation of Eq.(\ref{5}) over spatial coordinates $x, y, z$ with 
the  electrostatic potential $\varphi
\left(\textbf{r} ,t\right) $  determined as a 
function of coordinates $x$, $y$, $z$ as 
\begin{eqnarray}
&\displaystyle \varphi \left(x, y, z,
t\right) = \int\varphi \left(k_{x}, k_{y}, k_{z} ,t
\right)e^{ik_{x}x
+ik_{y}y+ik_{z}z} dk_{x}dk_{y}dk_{z},\label{6}
\end{eqnarray}
gives the equation for the separate spatial Fourier harmonic of $f_{\alpha}$ with  wave numbers $k_{x}, k_{y}, k_{z}$ 
\begin{eqnarray}
&\displaystyle \frac{\partial f_{\alpha}}{\partial t}+\left(i\left(k_{x}-V'_{0}tk_{y}\right)v_{\alpha x}+ik_{y}v_{\alpha y}+ik_{z}v_{\alpha z}\right)f_{\alpha}\left(\mathbf{v_{\alpha}}, \mathbf{k}, t \right)  + \omega_{c\alpha} v_{\alpha y}
\frac{\partial f_{\alpha}}{\partial v_{\alpha
x}}-\left(\omega_{c\alpha}+V'_{\alpha} \right) v_{\alpha
x}\frac{\partial f_{\alpha}}{\partial v_{\alpha y}} \nonumber
\\  &\displaystyle
=i\frac{e_{\alpha}}{m_{\alpha}}\varphi\left(\mathbf{k}, t \right)\left[\left(k_{x}-V'_{0}tk_{y}\right)  
\frac{\partial F_{0\alpha}}{\partial v_{\alpha x}}
+ik_{y}\frac{\partial F_{0\alpha}}{\partial v_{\alpha y}}
+ik_{z}\frac{\partial F_{0\alpha}}{\partial v_{\alpha
z}}\right].\label{7}
\end{eqnarray}
in which wave numbers $k_{x}, k_{y}, k_{z}$  and time change independently.
It follows from Eqs.(\ref{5}) and (\ref{7}), that the transformation of the Vlasov equation 
to convected-sheared coordinates (\ref{3}), (\ref{4}) converts the spatial inhomogeneity 
into the time inhomogeneity. That prevents the application of the spectral transforms in 
time to Eq.(\ref{7}) and to obtain the ordinary dispersion equations valid for any times.  
It reveals\cite{Mikhailenko-2011} that the ordinary modal solution to Eq.(\ref{7}), 
for which ordinary dispersion equation may be obtained, exists only at time 
$t\ll\left(V'_{0} \right) ^{-1}$; solution becomes non-modal\cite{Mikhailenko-2011} 
with time dependent frequency and growth rate for larger time.
With coordinates $\hat{x}$,~$\hat{y}$,~$\hat{z}$ of the laboratory set of references, 
transformation (\ref{6}) has a form
\begin{eqnarray}
&\displaystyle \varphi \left(\hat{x},\hat{y}, \hat{z},
t\right) = \int\varphi \left(k_{x}, k_{y}, k_{z} ,t
\right)e^{ik_{x}\hat{x}
+ik_{y}\left(\hat{y}-V'_{0}t\hat{x}\right) +ik_{z}\hat{z}} dk_{x}dk_{y}dk_{z}\nonumber
\\  &\displaystyle =
\int\varphi \left(k_{x}, k_{y}, k_{z} ,t
\right)e^{i\left( k_{x}-V'_{0}tk_{y}\right)\hat{x}
+ik_{y}\hat{y} +ik_{z}\hat{z}} dk_{x}dk_{y}dk_{z}.
\label{8}
\end{eqnarray}
It follows from Eq.(\ref{8}), that separate spatial Fourier mode in convected-sheared coordinates 
becomes a sheared mode with time dependent wave number $k_{x}-V'_{0}tk_{y}$ in the laboratory frame.

Usually, however, only the transformation to the convected coordinates (\ref{3}) in velocity space, 
without the transformation to the sheared coordinates (\ref{4}) in the configuration space, is used 
in the kinetic theory of plasma shear flows \cite{Artun, Sugama}. After such transformation
the linearized Vlasov equation becomes
\begin{eqnarray}
&\displaystyle \frac{\partial f_{\alpha}}{\partial
t} +V'_{0}\hat{x}\frac{\partial f_{\alpha}}{\partial
\hat{y}} + \hat{\mathbf{v}} \frac{\partial f_{\alpha}}
{\partial\hat{\mathbf{r}}} +\omega_{c\alpha} v_{\alpha y}
\frac{\partial f_{\alpha}}{\partial v_{\alpha
x}}-\left(\omega_{c\alpha}+V'_{\alpha} \right) v_{\alpha
x}\frac{\partial f_{\alpha}}{\partial v_{\alpha y}}\nonumber  \\
&\displaystyle = \frac{e_{\alpha}}{m_{\alpha}}\bigtriangledown
\varphi\left(\hat{\mathbf{r}},t\right)\frac{\partial
F_{0\alpha}}{\partial\hat{\mathbf{v}}}.\label{9}
\end{eqnarray}
Traditionally\cite{Artun, Sugama}, in proceeding with derivation of the governing equation for  
$f_{\alpha}$, the spatial Fourier transform in the laboratory configuration space,
\begin{eqnarray}
&\displaystyle \varphi \left(\hat{x},\hat{y}, \hat{z},
t\right) = \int\varphi \left(\hat{k}_{x}, \hat{k}_{y}, \hat{k}_{z} ,t
\right)e^{i\hat{k}_{x}\hat{x}
+i\hat{k}_{y}\hat{y} +i\hat{k}_{z}\hat{z}} d\hat{k}_{x}d\hat{k}_{y}d\hat{k}_{z}\label{10}
\end{eqnarray} 
for the electrostatic potential $\varphi$  and for $f_{\alpha}$ is adapted with assumption of
the "slow variation" of $V_{0}\left( x\right)$ with spatial coordinates. That 
gives the following equation for $f_{\alpha}\left(t, \hat{\mathbf{k}}, \mathbf{v} \right)$:
\begin{eqnarray}
&\displaystyle \frac{\partial f_{\alpha}}{\partial
t} -i\hat{k_{y}}V_{0}\left(\hat{x}\right)f_{\alpha}-i\left( \hat{\mathbf{k}}
\hat{\mathbf{v}}\right)f_{\alpha}+\omega_{c\alpha} v_{\alpha y}
\frac{\partial f_{\alpha}}{\partial v_{\alpha
x}}-\left(\omega_{c\alpha}+V'_{\alpha} \right) v_{\alpha
x}\frac{\partial f_{\alpha}}{\partial v_{\alpha y}} \nonumber  \\
&\displaystyle =-\frac{e_{\alpha}}{m_{\alpha}}\hat{\mathbf{k}}
\varphi\left(\hat{\mathbf{k}},t\right)\frac{\partial
F_{0\alpha}}{\partial\hat{\mathbf{v}}}.\label{11} 
\end{eqnarray}
In Eq.(\ref{10}), the flow velocity shear reveals in the formation of elliptical orbits of particles, 
with velocity coordinates $v_{\bot}$,~$\phi$ \cite{Shaing} 
\begin{eqnarray}
&\displaystyle v_{x}=v_{\bot}\cos \phi,\qquad v_{
y}=\sqrt{\eta} v_{\bot}\sin \phi, \qquad \phi=
\phi_{1}-\sqrt{\eta}\omega_{c}t, \qquad v_{z}=v_{z},\label{12}
\end{eqnarray}
where $\eta=1-V'_{0}/\omega_{c}$, and with modified gyro-frequency $\sqrt{\eta}\omega_{c}$.
That effect, however is negligible for $|V'_{0}|\ll \omega_{ci}$. 
Then, velocity shear is absorbed into the identical for both plasma species Doppler 
shifted frequency \cite{Artun, Sugama} $\hat{\omega}=\omega-k_{y}V_{0}\left(\hat{x}\right)$, 
and, in fact, is excluded from the subsequent analysis.

Let us analyse the results of the application of the Fourier transformation  (\ref{10}) to Eq.(\ref{9}) 
without the assumption of the "slow" spatial variation of the flow velocity. That 
gives the following differential equation in wave-number space:
\begin{eqnarray}
&\displaystyle \frac{\partial f_{\alpha}}{\partial
t} -V'_{0}\hat{k_{y}}\frac{\partial f_{\alpha}}{\partial
\hat{k_{x}}}-i\left( \hat{\mathbf{k}}
\hat{\mathbf{v}}\right)f_{\alpha}+\omega_{c\alpha} v_{\alpha y}
\frac{\partial f_{\alpha}}{\partial v_{\alpha
x}}-\left(\omega_{c\alpha}+V'_{\alpha} \right) v_{\alpha
x}\frac{\partial f_{\alpha}}{\partial v_{\alpha y}} \nonumber  \\
&\displaystyle =-\frac{e_{\alpha}}{m_{\alpha}}\hat{\mathbf{k}}
\varphi\left(\hat{\mathbf{k}},t\right)\frac{\partial
F_{0\alpha}}{\partial\hat{\mathbf{v}}}.\label{13} 
\end{eqnarray}
It is interesting to know the relation between solutions of Eqs. (\ref{7}) 
and (\ref{13}) for $f_{\alpha}$. For the receiving from Eq.(\ref{13}) the equation which 
couples $f_{\alpha}$ and $\varphi$ of the separate spatial Fourier mode, 
as it is in Eq.(\ref{5}), we have to exclude from Eq.(\ref{13}) the differential operator $-V'_{0}\hat{k}_{y}\frac{\partial f_{\alpha}}{\partial \hat{k}_{x}}$, due to which the 
Fourier mode of $f_{\alpha}$ appears to be coupled with all Fourier modes of the 
electrostatic potential and depends on the integral of $\varphi$ over wave-number space.
The characteristic equation 
\begin{eqnarray}
&\displaystyle
dt=-\frac{d\hat{k}_{x}}{V'_{0}\hat{k}_{y}}
\label{14} 
\end{eqnarray}
gives the solution $\hat{k}_{x}+V'_{0}t\hat{k}_{y}=K_{x}$, where $K_{x}$ 
as the integral of Eq.(\ref{13}) is time independent. It reveals that $f_{\alpha}
=f_{\alpha}\left(K_{x}, \hat{k}_{y}, \hat{k}_{z}, t\right)=f_{\alpha}\left(\hat{k}_{x}
+V'_{0}t\hat{k}_{y}, \hat{k}_{y}, \hat{k}_{z}, t\right)$, 
i.e. the wave number components $\hat{k}_{x}$ and $\hat{k}_{y}$ have to be changed 
in such a way that $\hat{k}_{x}+V'_{0}t\hat{k}_{y}$ leaves unchanged with time. 
The solution to Eq.(\ref{13})  for $f_{\alpha}$ can't be presented in the laboratory 
coordinates in a form, in which the time and spatial dependences are separable, 
as it is for the normal mode solutions of Eq. (\ref{11}) obtained with assumption 
of the "slow" spatial variation of the flow velocity\cite{Artun}.
If we use, however, $\hat{k}_{x}=K_{x}-V'_{0}t\hat{k}_{y}$ in Eqs.(\ref{10}) 
and (\ref{13}), we obtain for the electrostatic potential the presentation (\ref{8}), and  
we obtain Eq.(\ref{7}) for $f_{\alpha}$, with time independent 
$K_{x}=k_{x}$, $\hat{k}_{y}=k_{y}$, $\hat{k}_{z}=k_{z}$. The obtained results 
prove, that the solution of the Vlasov equation in the form of the separate 
Fourier harmonic with time independent wave numbers may be obtained only in 
convected-sheared coordinates. That solution reveals in the laboratory frame 
as a shearing mode (\ref{8}) with time dependent $x$-component of the  wave number. 

So, we have two procedures for the proper performing of the spatial Fourier 
transformation of the Vlasov equation for plasma shear flow, which give the same result.
The first one is to apply at first the transformation (\ref{2}) to convected-sheared coordinates, 
and then, to perform the Fourier transform of the Vlasov equation over spatial 
coordinates with time independent wave numbers $k_{x}$, $k_{y}$, $k_{z}$.  The second procedure is 
to come to these time independent wave numbers through the solution of the 
characteristic equation Eq.(\ref{14}), when the transformation to sheared 
coordinates in configuration space in the Vlasov equation does not perform. 
The sheared flow leads to the observed in the laboratory frame continuous inclining with time of the plane of 
constant phase the waves. That reveals in the observed in the laboratory frame 
the time dependence of the wave number $\hat{k}_{x}= K_{x}-V'_{0}\hat{k}_{y}t$ in the solution for the $f_{\alpha}$. 
The oversimplification of the problem, which resulted from the application 
of the assumption of "slow spatial variation of $V_{0}\left(x \right) $", leads 
to the overlooking of that principal effect of shear flow. It is obvious, 
that the time dependence in $K_{x}$ may be neglected only in the case 
of negligible velocity shear, or when the very short evolutionary time is considered. 
Really, for $\hat{k}_{y}\sim \hat{k}_{x}$ and $V'_{0}\simeq \gamma$ for time $t\gtrsim\gamma^{-1}$ the 
we have $\hat{k}_{y}V'_{0}t\gtrsim \hat{k}_{x}$ in the integral $K_{x}=\hat{k}_{x}+V'_{0}\hat{k}_{y}t$.
Therefore the assumption of "slow spatial variation of flow velocity" is not 
valid for the investigations of the effects of shear flow in real experiments, 
where observed velocity shearing rate may be of the order or above of the growth rate 
of the instability and the time of the observations is of the order of the inverse growth rate or longer.

With leading center coordinates $X, Y$, determined in convective-shearing coordinates by the relations
\begin{eqnarray}
&\displaystyle
x=X-\frac{v_{\bot}}{\sqrt{\eta}\omega_{c}}\sin
\phi, \qquad
y=Y+\frac{v_{\bot}}{\eta\omega_{c}}\cos
\phi +V'_{0}t\; \left(X-x\right), \nonumber \\
&\displaystyle
z_{1}=z-v_{z}t, \qquad \label{15}
\end{eqnarray}
the Vlasov equation (\ref{1}), in which species index is suppressed,  
transforms into the form \cite{Mikhailenko-2011} 
\begin{eqnarray}
&\displaystyle \frac{\partial F}{\partial
t}+\frac{e}{m\sqrt{\eta}\omega_{c}}
\left(\frac{\partial\varphi}{\partial X} \frac{\partial
F} {\partial Y}-\frac{\partial\varphi}{\partial
Y} \frac{\partial F} {\partial X}\right)\nonumber \\
&\displaystyle
+\frac{e}{m}\frac{\sqrt{\eta}\omega_{c}}{v_{\perp}}
\left(\frac{\partial\varphi}{\partial \phi_{1}} \frac{\partial
F} {\partial v_{\perp}}-\frac{\partial\varphi}{\partial
v_{\perp}}\frac{\partial F} {\partial \phi_{1}}\right)
-\frac{e}{m}\frac{\partial\varphi}{\partial
z_{1}} \frac{\partial F}{\partial v_{z}} =0,
\label{16}
\end{eqnarray}
in which any time dependent coefficients are absent. It follows from Eqs.(\ref{15}), that 
in shearing coordinates a particle gyro-motion is different from the ones in convective coordinates, 
determined by Eqs.(\ref{12}). Now it consists in the rotation with modified gyro-frequency and 
stretching of gyro-orbit along coordinate $y$ with velocity $-V'_{0}\left( x-X\right) $, 
which is negative for $x>X$ and is positive for $x<X$. With leading center coordinates (\ref{15})
the Fourier transform (\ref{6}) becomes
\begin{eqnarray}
&\displaystyle \varphi \left(x, y, z,
t\right) = \int\varphi \left(k_{x}, k_{y}, k_{z} ,t
\right)\exp\left[ ik_{x}\left( X-\frac{v_{\bot}}{\sqrt{\eta}\omega_{c}}\sin
\phi\right)\right. \nonumber \\
&\displaystyle \left.+ik_{y}\left( Y+\frac{v_{\bot}}{\eta\omega_{c}}\cos
\phi\right) +iV'_{0}t\frac{k_{y}v_{\bot}}{\sqrt{\eta}\omega_{c}}\sin
\phi +ik_{z}z\right]  dk_{x}dk_{y}dk_{z} \nonumber \\
&\displaystyle = \int\varphi \left(k_{x}, k_{y}, k_{z} ,t
\right)\exp\left[ ik_{x} X +ik_{y}Y+ik_{z}z\right.\nonumber \\
&\displaystyle \left.+i\frac{k_{y}v_{\bot}\cos
\phi}{\eta\omega_{c}}-i\frac{\left(k_{x}-V'_{0}tk_{y}\right) v_{\bot}\cos
\phi}{\sqrt{\eta}\omega_{c}}\right]dk_{x}dk_{y}dk_{z} = \nonumber  \\
&\displaystyle = \int\varphi \left(k_{x}, k_{y}, k_{z}, t \right)
\exp \left[ ik_{x}X_{i}+ik_{y}Y_{i}+ik_{z}z\right.  \nonumber  \\
&\displaystyle \left. -i\frac{\hat{k}_{\bot}\left(t\right)
v_{\bot}}{\sqrt{\eta}\omega_{ci}}\sin\left(
\phi_{1}-\sqrt{\eta}\omega_{ci}t -\theta
\left(t\right)\right)\right]dk_{x}dk_{y}dk_{z}
\label{17}
\end{eqnarray}
where
\begin{eqnarray}
&\displaystyle
\hat{k}^{2}_{\perp}\left(t\right)=\left(k_{x}-V'_{0}tk_{y}\right)^{2}+\frac{1}{\eta}k_{y}^{2},\label{18}
\end{eqnarray}
and $\tan \theta =k_{y}/\sqrt{\eta}(k_{x}-V'_{0}tk_{y})$. It follows from Eq.(\ref{17}) 
that finite Larmor radius effect of the interaction of the perturbation 
with time independent wave numbers $k_{x},\, k_{y},\, k_{z}$ with ion, Larmor 
orbit of which is observed in sheared coordinates as a spiral continuously stretched 
with time, appears identical analytically to the interaction of the perturbation with 
wave numbers $k_{x}-V'_{0}tk_{y}, \,\,k_{y}/\sqrt{\eta}, \,\,k_{z}$  with ion, which rotates on the 
elliptical orbit that is observed in the laboratory frame. The time dependence of 
the finite Larmor radius effect is the basic linear mechanism of the action of 
the velocity shear on waves and instabilities in plasma shear flow.

In what follows, we consider the equilibrium distribution function $F_{i0}$ as a
Maxwellian,
\begin{eqnarray}
& \displaystyle F_{0}=\frac{n_{0}\left(X
\right)}{\left(2\pi v^{2}_{T}\right)^{3/2}}
\exp\left(-\frac{v^{2}_{\bot}+v^{2}_{z}}{v^{2}_{T}} \right),
\label{19}
\end{eqnarray}
assuming the inhomogeneity of the density and temperature of plasma shear flow
species on coordinate $X$. In this paper we assume, that velocity shearing rate $V'_{0}$ is much less than 
the ion cyclotron frequency $\omega_{ci}$, and put $\eta=1$. The solution of the Vlasov equation for the
perturbation $f\left(t,k_{x},k_{y}, k_{z},
v_{\bot},\phi,v_{z},z_{1} \right)$ of the distribution function
$F$, $f=F-F_{0}$ with known $F_{0}$ is
\begin{eqnarray}
& \displaystyle f=
\frac{e}{m}\int\limits^{t}_{t_{o}}\left[\frac{1}{\omega_{c}}\frac{\partial\varphi}{
\partial Y}\frac{\partial F_{0}}{\partial X}-\frac{\omega_{c}}{v_{\bot}}
\frac{\partial\varphi}{\partial \phi_{1}} \frac{\partial
F_{0}}{\partial v_{\bot}} +\frac{\partial\varphi}{\partial z_{1}}
\frac{\partial F_{0}}{\partial v_{z}} \right] dt'. \label{20}
\end{eqnarray}

Using solution (\ref{20}) for all plasma species in Poisson equation
for the potential $\varphi\left(\textbf{r},t\right)$,
\begin{eqnarray}
&\displaystyle \vartriangle \varphi\left(\textbf{r},t\right)=
-4\pi\sum_{\alpha=i,e} e_{\alpha}\int f_{\alpha}\left(\textbf{v},
\textbf{r}, t \right)d\textbf {v}_{\alpha}, \label{21}
\end{eqnarray}
we obtain integral equation\cite{Mikhailenko-2011, Mikhailenko2010-1}, 
which governs the temporal evolution of the 
separate spatial Fourier harmonic of the electrostatic potential 
$\varphi\left(\mathbf{k},t\right)$ in plasma shear flow and 
is capable of handling linear as well as nonlinear\cite{Mikhailenko-2011} 
evolution of the electrostatic instabilities of plasma shear flows.

\section{HYDRODYNAMIC ION TEMPERATURE GRADIENT INSTABILITY}
In this section, we consider the temporal evolution of the hydrodynamic ion temperature 
gradient instability in plasma shear flow. By using the methodology stated in Sec.II, 
we obtain the integral equation for the electrostatic potential
$\Phi\left(\textbf{k}, t\right)=\varphi\left(\textbf{k},
t\right)\Theta\left(t-t_{0}\right)$, where $\Theta\left(t-t_{0}\right)$ is 
the unit-step Heaviside function (it is equal to zero for $t<t_{0}$ and equal to unity for $t\geq
t_{0}$)
\begin{eqnarray*}
&\displaystyle \int\limits^{t}_{t_{0}}dt_{1}
\frac{d}{dt_{1}}\left\lbrace \Phi\left(\mathbf{k},t_{1}\right)
\left[- \left(1+T\right)+A_{0i}\left(t, t_{1}\right) \right]\right\rbrace\nonumber
\\ &\displaystyle - \int\limits^{t}_{t_{0}}dt_{1}
\frac{d}{dt_{1}}\left\lbrace \Phi\left(\mathbf{k},t_{1}\right)
A_{0i}\left(t, t_{1}\right)\left( 1-e^{-\frac{1}{2}k^{2}_{z}v^{2}_{Ti}\left(t-t_{1}\right)^{2}}
\right) \right\rbrace  \nonumber \\ &\displaystyle 
+\int\limits^{t}_{t_{0}}dt_{1}
\Phi\left(\mathbf{k},t_{1}\right)
A_{0i}\left(t, t_{1}\right)
e^{-\frac{1}{2}k^{2}_{z}v^{2}_{Ti}
\left(t-t_{1}\right)^{2} } \nonumber
\\ &\displaystyle \times\left(ik_{y}v_{di}-i\omega_{Ti}-k^{2}_{z}v^{2}_{Ti}
\left(t-t_{1}\right)
-\frac{i}{2}\omega_{Ti}k^{2}_{z}v^{2}_{Ti}\left(t-t_{1}\right)^{2} \right)
\end{eqnarray*}
\begin{eqnarray}
&\displaystyle
+i\omega_{Ti}\int\limits^{t}_{t_{0}}dt_{1}
\Phi\left(\mathbf{k},t_{1}\right)A_{1i}\left(t, t_{1}\right)e^{-\frac{1}{2}k^{2}_{z}v^{2}_{Ti}
\left(t-t_{1}\right)^{2}} \nonumber
\\ &\displaystyle +T\int\limits^{t}_{t_{0}}dt_{1}
\left(\frac{d\Phi\left(\mathbf{k},t_{1}\right)}{dt_{1}}
+ik_{y}v_{de}\Phi\left(\mathbf{k},t_{1}\right)\right)e^{-\frac{1}{2}k^{2}_{z}v^{2}_{Te}
\left(t-t_{1}\right)^{2}}=0, \label{22}
\end{eqnarray}
where $T=T_{i}/T_{e}$, $\omega_{Ti}=k_{y}v_{di}d\ln T_{i}/d\ln n_{i}$, $v_{di}= \left(cT_{i}/eB\right)
d\ln n_{i0}/dx$ is the ion diamagnetic velocity,  
\begin{eqnarray}
&\displaystyle A_{0i}\left(t, t_{1}\right)=
I_{0}\left(\hat{k}_{\perp}\left(t\right)\hat{k}_{\perp}\left(t_{1}\right)
\rho^{2}_{i}\right)
e^{-\frac{1}{2}\rho^{2}_{i}\left(\hat{k}^{2}_{\perp}\left(t\right)+\hat{k}^{2}_{\perp}
\left(t_{1}\right)\right)},\label{23}
\end{eqnarray}
\begin{eqnarray}
&\displaystyle A_{1i}\left(t, t_{1}\right)=
e^{-\frac{1}{2}\rho^{2}_{i}\left(\hat{k}^{2}_{\perp}\left(t\right)
+\hat{k}^{2}_{\perp}\left(t_{1}\right)\right)}  \nonumber
\\ &\displaystyle \times \left[\left(1-\frac{\rho_{i}^{2}}{2}
\left(\hat{k}^{2}_{\perp}\left(t\right)
+\hat{k}^{2}_{\perp}\left(t_{1}\right)\right)\right) I_{0}\left(\hat{k}_{\perp}\left(t\right)\hat{k}_{\perp}\left(t_{1}\right)\rho^{2}_{i}\right)\right. 
\nonumber
\\ &\displaystyle \left.+ \rho^{2}_{i}\hat{k}_{\perp}\left(t\right)\hat{k}_{\perp}\left(t_{1}\right)
I_{1}\left(\hat{k}_{\perp}
\left(t\right)\hat{k}_{\perp}\left(t_{1}\right)\rho^{2}_{i}\right) \right].\label{24}
\end{eqnarray}
Eq.(\ref{22}) is the extension of the Eq.(25) in Ref.\cite{Mikhailenko-2011} 
onto the accounting for the inhomogeneity of the ion temperature.
Naturally, it is not possible to obtain explicit single analytical presentation 
of the solution to Eq.(\ref{22}), which is valid for any desired time. The exceptional advantage 
of the application of the non-modal approach, which 
uses the wavenumber-time variables, is a possibility to perform the analysis 
of the electrostatic potential evolution at any finite time domain and with an 
arbitrary initial time $t_{0}$. In our analysis, 
initiated in \cite{Mikhailenko-2011}, we distinguish different characteristic times 
of the evolution process of the long wavelength perturbations with $\hat{k}_{\bot}\left(t_{0} \right)\rho_{i}\ll 1 $: $t_{0}$ is the time of the perturbation origin; $t_{(0)}=0$ is the time of the electric 
field (shear flow) origin; times $t_{(1)}=\left( V'_{0}\right)^{-1}$, $t_{(2)}=t_{s}
=\left( V'_{0}k_{y}\rho_{i}\right)^{-1}$ define different stages of the non-modal 
evolution of the potential under the action of the shear flow. In these time intervals 
different tasks in the solving of Eq.(\ref{22}) arise (as it was with Eq.(25) 
in Ref.\cite{Mikhailenko-2011}). We will derive the linear solution of Eq.(\ref{22}) 
at time $\left( V'_{0}\right)^{-1} \gg t \gg t_{0} $ of the modal instability development,
at which shear flow effects are underdeveloped. Then, we will derive the solution at time 
$t\gg \left( V'_{0}\right)^{-1}$, at which the non-modal effects of the shear flow become
pronounced. And, at last, we will derive solution at the final stage of the perturbation 
evolution at time $t\gg t_{s}=\left( V'_{0}k_{y}\rho_{i}\right)^{-1}\gg \left( V'_{0}\right)^{-1}$, at 
which even initially long wavelength perturbations with $\hat{k}_{\perp}\rho_{i}<1$ 
at time $t\ll t_{s}$, becomes the short wavelength ones with $\hat{k}_{\perp}\rho_{i}\gg 1$ 
at $t\gg t_{s}$. 
\begin{figure}
\includegraphics[width=0.9\textwidth]{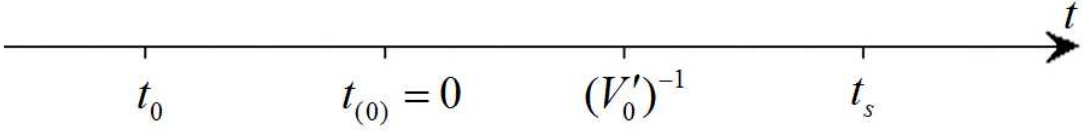}
\caption{\label{fig1}The sequence of the characteristic times for Eq.(\ref{22}) for 
the long wavelength perturbations with $\hat{k}_{\bot} \left(t_{0} \right)\rho_{i}< 1 $.}
\end{figure}

The long-wavelength drift perturbations with $\hat{k}_{\bot}\rho_{i}<1$ at time $t=0$,  will be the
long wavelength ones at time $t< t_{s}=\left( V'_{0}k_{y}\rho_{i}\right)^{-1}$. For this limit, 
we use the approximations, 
\begin{eqnarray}
&\displaystyle
A_{0i}\left(t, t_{1}\right)\approx 1-\frac{\rho_{i}^{2}}{2}\left(\hat{k}^{2}_{\perp}\left(t\right)
+\hat{k}^{2}_{\perp}\left(t_{1}\right)\right) \nonumber \\ &\displaystyle
\approx b_{i}+\rho^{2}_{i}\left(k_{x}k_{y}V'_{0}\left(t+
t_{1}\right)-\frac{1}{2}k^{2}_{y}
\left(V'_{0}\right)^{2}\left(t^{2}+t^{2}_{1}\right)\right)
\Theta\left(t_{1}\right),\label{25}
\end{eqnarray}
\begin{eqnarray}
&\displaystyle
A_{1i}\left(t, t_{1}\right)\approx 1-\rho_{i}^{2}\left(\hat{k}^{2}_{\perp}\left(t\right)
+\hat{k}^{2}_{\perp}\left(t_{1}\right)\right) \nonumber \\ &\displaystyle
\approx b_{i1}+\rho^{2}_{i}\left(2k_{x}k_{y}V'_{0}\left(t+
t_{1}\right)-k^{2}_{y}
\left(V'_{0}\right)^{2}\left(t^{2}+t^{2}_{1}\right)\right)\Theta\left(t_{1}\right),\label{26}
\end{eqnarray}
where $b_{i}=1-k^{2}_{\perp}\rho^{2}_{i}, \,\,b_{i1}=1-2k^{2}_{\perp}\rho^{2}_{i}, \,\, k^{2}_{\perp}=k^{2}_{x}+k^{2}_{y}$, and $\Theta\left(t\right)$
indicates that the shear flow emerges at $t=0$.
In Eq.(\ref{22}), we use the approximations
$1-\exp\left(-(1/2)k_{z}^{2}v_{Ti}^{2}\left(t-t_{1} \right)^{2} \right)
\simeq -(1/2)k_{z}^{2}v_{Ti}^{2}\left(t-t_{1} \right)^{2} $, that corresponds to the weak 
ion Landau damping. Assuming that the 
magnitude of the initial potential $\Phi\left(\mathbf{k}, t= t_{0}\right)$ is 
negligibly small, the integral
\begin{eqnarray}
&\displaystyle \int\limits^{t}_{t_{0}}dt_{1}
\frac{d}{dt_{1}}\left( \Phi\left(\mathbf{k},t_{1}\right)\left(1- 
e^{-\frac{1}{2}k^{2}_{z}v^{2}_{Ti}\left(t-t_{1}\right)^{2}}\right)\right)\nonumber
\\ &\displaystyle   \simeq \int\limits^{t}_{t_{0}}dt_{1}
\frac{d}{dt_{1}}\left(\Phi\left(\mathbf{k},t_{1}\right)\left(\frac{1}{2}
k^{2}_{z}v^{2}_{Ti}\left(t-t_{1}\right)^{2}\right)\right)\nonumber
\\ &\displaystyle =\frac{1}{2}k^{2}_{z}v^{2}_{Ti}\left(t-t_{0}\right)^{2}
\Phi\left(\mathbf{k},t_{0}\right) \label{27}
\end{eqnarray} 
may be ignored in Eq.(\ref{22}). This approximation is reasonable only for the 
exponentially growing, faster than $\left(t-t_{0}\right)$ or $\left(t-t_{0}\right)^{2}$,
potential $\Phi\left(\mathbf{k},t\right)$; then $\Phi\left(\mathbf{k},t_{0}\right)
\left(t-t_{0} \right)$ and $\left( d\Phi\left(\mathbf{k},t\right)/dt\right)|_{t=t_{0}}
\cdot \left(t-t_{0} \right)^{2}$ are exponentially small with respect to  
$\Phi\left(\mathbf{k},t\right)$. We assume that $k^{2}_{z}v^{2}_{Te} \left(t-t_{1}\right)^{2}\gg 1$, 
that corresponds to adiabatic electrons, and neglect exponentially small electron terms in 
Eq.(\ref{22}). In result, we obtain the simplified integral 
equation, which describes the linear evolution of the electrostatic potential at time $t\ll t_{s}$,
\begin{eqnarray*}
&\displaystyle \left(T+k^{2}_{\perp}\rho^{2}_{i}\right)\int\limits^{t}_{t_{0}}dt_{1}
\frac{d^{3}\Psi\left(\mathbf{k},t_{1}\right)}{dt^{3}_{1}}  \nonumber
\\ &\displaystyle -\int\limits^{t}_{t_{0}}dt_{1}
\frac{d^{2}\Psi\left(\mathbf{k},t_{1}\right)}{dt^{2}_{1}}\left[b_{i}\left(ik_{y}v_{di}
-k^{2}_{z}v^{2}_{Ti}\left(t-t_{1}\right)\right.\right.\nonumber
\\ &\displaystyle\left.\left.-\frac{i}{2}\omega_{Ti}k^{2}_{z}v^{2}_{Ti}
\left(t-t_{1}\right)^{2} \right)-i\omega_{Ti}k_{\perp}^{2}\rho_{i}^{2}\right]
\end{eqnarray*} 
\begin{eqnarray} &\displaystyle = \rho^{2}_{i}\int\limits^{t}_{t_{0}}dt_{1}\frac{d}{dt_{1} }\left[\frac{d^{2}\Psi\left(\mathbf{k},t_{1}\right)}{dt^{2}_{1}}\left( k_{x}k_{y}V'_{0}
\left(t+t_{1}\right) - \frac{1}{2}k_{y}^{2}\left( V'_{0}\right)^{2}\left(t^{2}+t^{2}_{1}\right)\right)\Theta\left(t_{1} \right)  \right]
\nonumber \\ &\displaystyle +
\rho^{2}_{i}\int\limits^{t}_{t_{0}}dt_{1} 
\frac{d^{2}\Psi\left(\mathbf{k},t_{1}\right)}{dt^{2}_{1}}\left(k_{x}k_{y}V'_{0}\left(t+
t_{1}\right)-\frac{1}{2}k^{2}_{y}
\left(V'_{0}\right)^{2}\left(t^{2}+t^{2}_{1}\right)\right)
\Theta\left(t_{1}\right)  \nonumber \\ &\displaystyle 
\times\left[ik_{y}v_{di}+ i\omega_{Ti} -k^{2}_{z}v^{2}_{Ti}\left(t-t_{1}\right)-
i\frac{\omega_{Ti}}{2}k^{2}_{z}v^{2}_{Ti}\left(t-t_{1}\right)^{2}\right].   \label{28}
\end{eqnarray} 
where new variable $\Psi$, determined as $d^{2}\Psi/dt^{2}=\Phi$, is introduced. 
Now we obtain the solution to Eq.(\ref{28}) in the limit of large value of the 
parameter $\eta_{i} \gg 1$, at which long wavelength with $k_{\perp}\rho_{i}<1$ hydrodynamic  ion 
temperature gradient instability is developed in a shearless plasma. By integration 
in parts of Eq.(\ref{28}) and neglecting the initial 
values of $d\Psi\left(\mathbf{k},t\right)/dt$ and $\Psi\left(\mathbf{k},t\right)$
at $t=t_{0}$ with  assumption of the exponential growth for  $\Psi\left(\mathbf{k},t\right)$ at 
time $t\gg t_{0}$, we obtain in the zero approximation the equation
\begin{eqnarray}
&\displaystyle \int\limits^{t}_{t_{0}}dt_{1}\left[ \left(T+k^{2}_{\perp}\rho^{2}_{i}\right)\frac{d^{3}\Psi\left(\mathbf{k},t_{1}\right)}{dt^{3}_{1}}
+i\omega_{Ti}k^{2}_{z}v_{Ti}b_{i}\Psi\left(\mathbf{k},t_{1}\right) \right] =0, \label{29}
\end{eqnarray}
in which the dominant terms in the left-hand side of Eq.(\ref{28}) are retained, and the right-hand side, 
proportional to $k_{y}^{2}\rho^{2}_{i}\ll 1$, is neglected. The solution to Eq.(\ref{29}) is 
$\Psi\left(\mathbf{k},t\right)=C\exp\left(-i\omega\left(\mathbf{k}\right)t\right)$, where the frequency $\omega\left(\mathbf{k}\right)$ is determined by the known equation
\begin{eqnarray}
&\displaystyle \omega^{3}\left(\mathbf{k}\right)=-\omega_{Ti}k^{2}_{z}v^{2}_{Ti} 
\frac{\left(1- k^{2}_{\perp}\rho^{2}_{i}\right)}{\left(T+k^{2}_{\perp}\rho^{2}_{i}\right)}.\label{30}
\end{eqnarray} 
The accounting for the small right-hand side of Eq.(\ref{28}) will modify that solution. 
We seek in the next approximation the solution for $\Psi\left(\mathbf{k},t\right)$ in the form
\begin{eqnarray}
&\displaystyle
\Psi\left(\mathbf{k},t\right)=C\exp\left(-i\omega\left(\mathbf{k}\right)t+\sigma
\left(\mathbf{k}, t\right)\right),\label{31}
\end{eqnarray}
where the higher order correction, $\sigma\left(\mathbf{k}, t\right)$, we find by use the 
procedure of the successive approximations\cite{Mikhailenko-2011}, which gives the following 
equation for $\sigma\left(\mathbf{k}, t\right)$:
\begin{eqnarray}
&\displaystyle
3\left(T+k^{2}_{\perp}\rho^{2}_{i}\right)\omega^{2}\left(\mathbf{k}\right)\int\limits^{t}_{t_{0}}dt_{1}
\Psi\left(\mathbf{k},t_{1}\right)\frac{d\sigma}{dt_{1}}\nonumber
\\ &\displaystyle = -\rho^{2}_{i}\int\limits^{t}_{t_{0}}dt_{1}\frac{d}{dt_{1} }\left[\frac{d^{2}\Psi\left(\mathbf{k},t_{1}\right)}{dt^{2}_{1}}\left( k_{x}k_{y}V'_{0}
\left(t+t_{1}\right) - \frac{1}{2}k_{y}^{2}\left( V'_{0}\right)^{2}\left(t^{2}+t^{2}_{1}\right)\right)\Theta\left(t_{1} \right)  \right]\nonumber
\\ &\displaystyle +i\frac{\omega_{Ti}}{2}\rho^{2}_{i}\int\limits^{t}_{t_{0}}dt_{1} 
\frac{d^{2}\Psi\left(\mathbf{k},t_{1}\right)}{dt^{2}_{1}}\left(k_{x}k_{y}V'_{0}\left(t+
t_{1}\right)-\frac{1}{2}k^{2}_{y}
\left(V'_{0}\right)^{2}\left(t^{2}+t^{2}_{1}\right)\right)
\Theta\left(t_{1}\right)  \nonumber \\ &\displaystyle 
\times k^{2}_{z}v^{2}_{Ti}\left(t-t_{1}\right)^{2},\label{32}
\end{eqnarray}
where $\sigma\left(\mathbf{k}, t\right)$ is neglected in $\Psi\left(\mathbf{k},t_{1}\right)$ 
in the right-hand side. The solution to Eq.(\ref{32}) is obtained straightforwardly and is equal to
\begin{eqnarray}
&\displaystyle
\sigma\left(\mathbf{k}, t\right)=-\frac{1}{3}k^{2}_{y}\rho^{2}_{i}\lambda \left(V'_{0}t\right)^{2}
+\frac{i}{9}k^{2}_{y}\rho^{2}_{i}\omega\left(\mathbf{k}\right)t\lambda \left(V'_{0}t\right)^{2},
\label{33}
\end{eqnarray}
where 
\begin{eqnarray*}
&\displaystyle \lambda =\frac{1-T-2k^{2}_{\perp}\rho^{2}_{i}}{\left(1-k^{2}_{\perp}\rho^{2}_{i}\right)
\left(T+k^{2}_{\perp}\rho^{2}_{i} \right)}.
\end{eqnarray*}
This result displays, that linear effect of the shear flow at time $t_{s}\gg t\gg 
\left(V'_{0}\right)^{-1}$ 
consists in the non-modal decrease of the frequency and the growth rate with time.
Qualitatively the same result was obtained for the drift kinetic instability in Ref.\cite{Mikhailenko-2011}. 

It may be anticipated that the most substantial effect of the instability suppression 
will be attained at time $t$ approaching $t_{s}$ and that suppression will continue at $t>t_{s}$. 
For that time, however, solution to Eq.(\ref{22}) may be obtained only numerically. 
For time $t\gg t_{s}$,  we have $\hat{k}_{\perp}\left(t\right)\rho_{i}\approx k_{y}
V'_{0}t\rho_{i}=t/t_{s}\gg 1$ and small parameter $t_{s}/t\sim t_{s}/t_{0}\ll 1$ will 
appear in Eq.(\ref{22}), that admits the receiving of the analytical solution\cite{Mikhailenko-2011}
 to Eq.(\ref{22}). 
For that time
\begin{eqnarray}
&\displaystyle
A_{0i}\left(t, t_{1}\right)\approx \frac{t_{s}}{\sqrt{2\pi t t_{1}}}
e^{-\frac{1}{2t^{2}_{s}}\left(t-t_{1}\right)^{2}},
\label{34}
\end{eqnarray}
\begin{eqnarray}
&\displaystyle
A_{1i}\left(t, t_{1}\right)\approx \frac{t_{s}}{\sqrt{2\pi t t_{1}}}
e^{-\frac{1}{2t^{2}_{s}}\left(t-t_{1}\right)^{2}}\left(1-\frac{1}{2t^{2}_{s}}\left(t-t_{1}\right)^{2} \right). 
\label{35}
\end{eqnarray}

\begin{figure}
\includegraphics[width=0.8\textwidth]{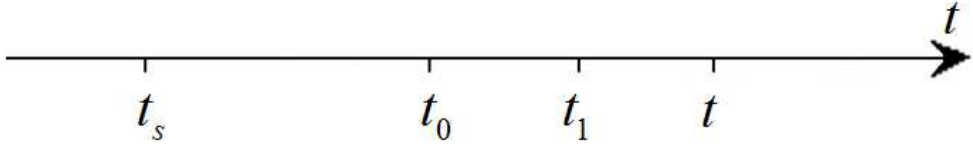}
\caption{\label{fig2}The domain of the integration $[t_{0}, t]$ over time in Eq.(\ref{36}) 
for the perturbations with $\hat{k}_{\bot}\left(t_{0} \right)\rho_{i}\gg 1$.}
\end{figure}

For the application the asymptotics (\ref{34}), (\ref{35}) for time $t_{1}$ in the whole 
interval $t>t_{1}>t_{0}$, we  have to consider in integral equation (\ref{32}) the initial 
time $t_{0}>t_{s}$. At that case the condition 
$\varphi \left(\mathbf{k}, t\right)=0$,  for $\left( V'_{0}\right) ^{-1}<t_{0}$ which was 
used for the function $\Phi \left(\mathbf{k}, t\right)$ in Eq.(\ref{22}), in which time 
$t_{0}$ was considered as the time preceding the development of the modal instability and 
appearance of the shear flow, is not applicable now, and we have to restore the equation 
for $\varphi \left(\mathbf{k}, t\right)$, obtained in Ref.\cite{Mikhailenko-2011}. It is important to note 
also, that for the decaying with time solution for $\varphi \left(\mathbf{k}, t\right)$, 
the approximation of the neglecting by initial value $\varphi \left(\mathbf{k}, t_{0}\right)$ 
used in Eq.(\ref{28}) is not justified and we have to retain these terms in equation considered.
We obtain finally the following integral equation:
\begin{eqnarray}
&\displaystyle \left(1+T \right)\int\limits^{t}_{t_{0}}dt_{1}\frac{d\varphi 
\left(\mathbf{k}, t\right)}{dt_{t}}= \int\limits^{t}_{t_{0}}dt_{1}\frac{d}
{dt_{t}}\left[\varphi \left(\mathbf{k}, t\right)\frac{t_{s}}
{\sqrt{2\pi t t_{1}}} e^{-\frac{\kappa^{2}_{i}}{2}\left(t-t_{1}\right)^{2}}\right]  
\nonumber \\
&\displaystyle +\int\limits^{t}_{t_{0}}dt_{1}\varphi \left(\mathbf{k}, t\right)
\frac{t_{s}}{\sqrt{2\pi t t_{1}}}
e^{-\frac{\kappa^{2}_{i}}{2}\left(t-t_{1}\right)^{2}}  \nonumber \\
&\displaystyle \times \left(ik_{y}v_{di}-\frac{i}{2}\omega_{Ti}\kappa^{2}_{i}
\left(t-t_{1}\right)^{2}-k^{2}_{z}v^{2}_{Ti}\left(t-t_{1}\right) \right) \nonumber \\
&\displaystyle +\varphi \left(\mathbf{k}, t_{0}\right)\frac{t_{s}}{t_{0}\sqrt{2\pi}}\left( 
e^{-\frac{1}{2t^{2}_{s}}\left(t-t_{0}\right)^{2}}- 1
\right),\label{36}
\end{eqnarray} 
where $\kappa^{2}_{i}=t^{-2}_{s}+k^{2}_{z}v^{2}_{Ti}$. At $t\gg t_{s}$,  the right-hand side of 
Eq.(\ref{36}) is proportional to small parameter $t_{s}/t\sim t_{s}/t_{0}$. In zero order in that 
parameter we have the equation $\int\limits^{t}_{t_{0}}dt_{1}\left( d\varphi \left(\mathbf{k}, t_{1}\right)/dt_{1}\right) =0$ 
with solution $\varphi \left(\mathbf{k}, t\right)=const$. The solution, which accounted 
for the terms of the order of $t_{s}/t$ in Eq.(\ref{36}), we find in the form $\varphi \left(\mathbf{k}, t\right)=C\exp\left(\sigma \left(\mathbf{k}, t\right)\right) $, where
$\sigma \left(\mathbf{k}, t\right) =O\left(t_{s}/t \right)$. Omitting $\sigma \left(\mathbf{k}, t\right)$ 
in the right-hand part of Eq.(\ref{36}), we obtain the approximate solution for $\varphi \left(\mathbf{k}, t\right)$,
\begin{eqnarray}
&\displaystyle
\varphi \left(\mathbf{k}, t\right)=\varphi_{0}\exp \left[\frac{t_{s}}{t\left(1+T\right)}
\left(i\frac{k_{y}v_{di}}{2\kappa_{i}}-i\frac{\omega_{Ti}}{4\kappa_{i}}+\sqrt{\frac{2}{\pi}}
\frac{k^{2}_{z}v^{2}_{Ti}}{\kappa^{2}_{i}} \right)  \right], 
\label{37}
\end{eqnarray}
which is qualitatively the same as it was obtained in Ref.\cite{Mikhailenko-2011} for the kinetic drift instability for large time, $t \gg t_{s}$ and is far from ordinary modal solution.

\section*{IV.NON-MODAL KINETIC EVOLUTION \\ OF THE HYDRODYNAMIC 
RESISTIVE  DRIFT INSTABILITY}

In this section, we consider the temporal evolution of the resistive drift 
instability in plasma with comparable temperatures of ions and electrons. 
In convective electron guiding center coordinates, linear perturbation $f_{e}$ of 
the steady state electron distribution function $F_{0e}$ is governed by linearized Vlasov 
equation with number-conserving Bhatnagar-Gross-Krook (BGK)
collision term\cite{Bhatnagar}
\begin{eqnarray}
&\displaystyle \frac{\partial f_{e}}{\partial
t}=\frac{e}{m_{e}\omega_{ce}}
\frac{\partial\varphi}{\partial Y_{e}} \frac{\partial F_{0e}}
{\partial X_{e}} -\frac{e}{m_{e}}\frac{\omega_{ce}}{v_{\perp}}
\frac{\partial\varphi}{\partial \phi_{1}} \frac{\partial F_{0e}}
{\partial v_{\perp}} +\frac{e}{m_{e}}\frac{\partial\varphi}{\partial
z} \frac{\partial F_{0e}}{\partial v_{z}} -\nu_{e}\left(f_{e}-\frac{n_{e1}\left(\textbf{r}, t\right)}{n_{e0}}F_{0e}\right).\label{38}
\end{eqnarray}
In Eq.(\ref{38}) $n_{e1}$ is the perturbed electron density, and $X_{e}$ is electron 
leading center coordinate. The solution to Eq.(\ref{38}) is 
\begin{eqnarray}
& \displaystyle f_{e}=
\frac{e}{m_{e}}\int\limits^{t}_{t_{o}}e^{-\nu_{e}\left(t-t_{1}\right)}\left[\frac{1}{\omega_{ce}}\frac{\partial\varphi}{
\partial Y_{e}}\frac{\partial F_{e0}}{\partial X_{e}}-\frac{\omega_{ce}}{v_{\bot}}
\frac{\partial\varphi}{\partial \phi_{1}} \frac{\partial
F_{e0}}{\partial v_{\bot}} +\frac{\partial\varphi}{\partial z_{1}}
\frac{\partial F_{e0}}{\partial v_{z}} \right] dt_{1} \nonumber \\
&\displaystyle +\nu_{e}\int\limits^{t}_{t_{o}}e^{-\nu_{e}\left(t-t_{1}\right)}\frac{n_{e1}\left(\textbf{r}, t_{1}\right)}{n_{e0}}F_{0e}dt_{1}. \label{39}
\end{eqnarray}
Using solution (\ref{39}) with Maxwellian distribution $F_{0e}$, (\ref{19}), and solution (\ref{20})
for the perturbation of the ion distribution function in Poisson equation (\ref{21}), we obtain 
in the quasineutrality approximation the following equation for the potential 
$\Phi\left(\mathbf{k},t\right)=
\varphi\left(\mathbf{k},t\right)\Theta \left(t-t_{0}\right)$ for low frequency, 
$d\varphi/dt \ll \omega_{ci}\varphi$, perturbations of drift type:
\begin{eqnarray}
&\displaystyle \int\limits^{t}_{t_{0}}dt_{1}\left\lbrace\left(1+T
\right)\frac{d}{dt_{1}}\Phi\left(\mathbf{k},t_{1}\right) -\frac{d}{dt_{1}}
\left(\Phi\left(\mathbf{k},t_{1}\right)
A_{0i}\left(t, t_{1} \right) \right)e^{-\frac{1}{2}k^{2}_{z}v^{2}_{Ti}
\left(t-t_{1}\right)^{2}}\right\rbrace \nonumber
\\ &\displaystyle
=i\int\limits^{t}_{t_{0}}dt_{1}
\Phi\left(\mathbf{k},t_{1}\right)k_{y}v_{di}A_{0i}\left(t, t_{1} \right) 
+ TG_{e}\left(\mathbf{k},t \right) ,\label{40}
\end{eqnarray}
where $T=T_{i}/T_{e}$ and $\Theta\left(t-t_{0}\right)$ is the unit-step Heaviside function. $G_{e}\left(\mathbf{k},t\right)$ determines the nonadiabatic part of the electron density perturbation,
\begin{eqnarray}
&\displaystyle n_{e1}\left(\mathbf{k}, t\right)=-\frac{en_{e0}}{T_{e}}
\Phi\left(\mathbf{k},t\right)+\frac{en_{e0}}{T_{e}}G_{e}\left(\mathbf{k},t\right).\label{41}
\end{eqnarray}
and is equal to
\begin{eqnarray}
&\displaystyle G_{e}\left(\mathbf{k},t \right)=\int\limits^{t}_{t_{0}}dt_{1}\left(
\frac{d}{dt_{1}}\left(\Phi\left(\mathbf{k},t_{1}\right)e^{-\nu_{e}\left(t-t_{1} \right)} \right)+ik_{y}v_{de}\Phi\left(\mathbf{k},t_{1}\right)e^{-\nu_{e}\left(t-t_{1}\right)}\right)e^{-\frac{1}{2}k^{2}_{z}v^{2}_{Te}\left(t-t_{1}\right)^{2}} \nonumber
\\ &\displaystyle   +4\pi\nu_{e}\lambda^{2}_{De}\int\limits^{t}_{t_{0}}dt_{1}n_{e1}
\left(\mathbf{k},t_{1}\right)e^{-\nu_{e}\left(t-t_{1}\right)
-\frac{1}{2}k^{2}_{z}v^{2}_{Te}\left(t-t_{1}\right)^{2}}.\label{42}
\end{eqnarray}

Now we apply the methodology developed in Ref.\cite{Mikhailenko-2011} to calculate 
the approximate solution to system (\ref{40}), (\ref{42}) for long wavelength 
perturbations with $\hat{k}_{\perp}\left(t\right)\rho_{i}<1$ 
for times limited by the condition $\left( V'_{0}\right)^{-1}<t<t_{s}$.
Making use the approximation (\ref{23}) we present Eq.(\ref{40}) in the form
\begin{eqnarray*}
&\displaystyle \int\limits^{t}_{t_{0}}dt_{1} \left(\frac{d
\Phi\left(\mathbf{k},t_{1}\right)}{dt_{1}}+i\omega\left(\mathbf{k}\right)
\Phi\left(\mathbf{k},t_{1}\right)\right)\nonumber
\\ &\displaystyle=-\frac{b_{i}}{T+k^{2}_{\perp}\rho^{2}_{i}}\int\limits^{t}_{t_{0}}dt_{1}\left(\frac{d
\Phi\left(\mathbf{k},t_{1}\right)}{dt_{1}}+ik_{y}v_{di}
\Phi\left(\mathbf{k},t_{1}\right)\right)\left(1-\exp\left( -\frac{1}{2}k^{2}_{z}v^{2}_{Ti}
\left(t-t_{1}\right)^{2}\right) \right)
\end{eqnarray*}
\begin{eqnarray}
&\displaystyle+\frac{1}{T+k^{2}_{\perp}\rho^{2}_{i}}\int\limits^{t}_{0}dt_{1}\left(\frac{d
\Phi\left(\mathbf{k},t_{1}\right)}{dt_{1}}+ik_{y}v_{di}
\Phi\left(\mathbf{k},t_{1}\right)\right)\left(\frac{k_{x}}{k_{y}}\frac{\left(t+t_{1}\right)}
{V'_{0}t^{2}_{s}}-\frac{\left(t^{2}+t^{2}_{1}\right)}{2t^{2}_{s}}\right)\nonumber \\
&\displaystyle+\frac{1}{T+k^{2}_{\perp}\rho^{2}_{i}}\int\limits^{t}_{0}dt_{1}
\Phi\left(\mathbf{k},t_{1}\right) \frac{1}{V'_{0}t^{2}_{s}}\left(\frac{k_{x}}{k_{y}}
-V'_{0}t\right)+\frac{T}{T+k^{2}_{\perp}\rho^{2}_{i}}G_{e}\left(\mathbf{k},t \right),\label{43}
\end{eqnarray}
where  $\omega\left(\mathbf{k}\right)$ is
\begin{eqnarray}
&\displaystyle
\omega\left(\mathbf{k}\right)=\frac{b_{i}}{1+k^{2}_{\perp}\rho^{2}_{s}}k_{y}v_{de},
\label{44}
\end{eqnarray}
$\rho_{s}$ is the ion thermal Larmor radius with electron temperature, and $v_{de}$ 
is the electron diamagnetic velocity. The first term in the right-hand side of Eq.(\ref{43})
determines the ion Landau damping; this
term is the same as in plasma without shear flow. The next three
terms originate from shear flow and determine the corrections to
the frequency and growth rate of the ordinary hydrodynamic resistive drift instability, which 
are provided by shear flow. The right-hand side of Eq.(\ref{43}) is small for
$\left(V'_{0}\right)^{-1}<t<t_{s}$, and because of the weakness of ion Landau
damping. We equate the right-hand side of  Eq.(\ref{43}) to zero and obtain in 
the lowest order the modal solution,
$\Phi\left(\mathbf{k},t\right)=\Phi_{0}\exp \left(-i\omega\left(\mathbf{k}\right) t\right)$. Making use 
this solution in Eq.(\ref{42}) to evaluate $G_{e}\left(\mathbf{k}, t\right)$, we obtain
\begin{eqnarray}
&\displaystyle G_{e}\left(\mathbf{k}, t\right)=-i\frac{\frac{\left( \omega\left(\mathbf{k}\right)
-k_{y}v_{de}\right) }{k_{z}v_{Te}}\sqrt{\frac{\pi}{2}}W\left(\frac{\omega
\left(\mathbf{k}\right)+i\nu_{e}}{\sqrt{2}k_{z}v_{Te}} \right)}{1-\frac{\nu_{e}}{k_{z}v_{Te}}\sqrt{\frac{\pi}{2}}W\left(\frac{\omega
\left(\mathbf{k}\right)+i\nu_{e}}{\sqrt{2}k_{z}v_{Te}} \right)}\Phi\left(\mathbf{k},t\right),
\label{45}
\end{eqnarray}
where  $W\left(z\right)=e^{ - z^{2}}\left(1 +\left(2i / \sqrt {\pi }
\right)\int\limits_{0}^{z} e^{t^{2}}dt \right)$ is plasma dispersion function. 
Accounting for the small right-hand side of Eq.(\ref{43}), we seek in the next approximation the 
solution in the form
\begin{eqnarray}
&\displaystyle
\Phi\left(\mathbf{k},t\right)=\Phi_{0}\exp\left(-i\omega\left(\mathbf{k}\right)t+\sigma
\left(\mathbf{k}, t\right)\right).\label{46}
\end{eqnarray}
Assuming that all terms at the right-hand side of 
Eq.(\ref{41}) are of the same order,  by inserting Eq.(\ref{46}) into Eq.(\ref{43}) and neglecting  derivative $d\sigma \left(\mathbf{k}, t\right)/dt$  in the right-hand side of
Eq.(\ref{43}), the equation  for $\sigma\left(\mathbf{k}, t\right)$ is found
\begin{eqnarray}
&\displaystyle \int\limits^{t}_{t_{0}\rightarrow
-\infty}dt_{1}\Phi\left(\mathbf{k},t_{1}\right)\left[\frac{d\sigma
\left(\mathbf{k}, t_{1}\right)}{dt_{1}}-i\delta\omega\left(\mathbf{k}\right)\right.\nonumber
\\ &\displaystyle\left.-\frac{\Theta\left(t_{1}\right)} {\left( T+k^{2}_{\perp}\rho^{2}_{i}\right) t^{2}_{s}}
\left(i\omega\left(\mathbf{k}\right) t^{2}_{1}\frac{\left(1+T\right)}{b_{i}}-2t_{1} \left(1+\frac{T+k^{2}_{\perp}\rho^{2}_{i}}{2b_{i}}\right)\right)\right]=0,\label{47}
\end{eqnarray}
in which was  assumed that $t_{0}\rightarrow -\infty$. In this equation, $\delta\omega\left(\mathbf{k}\right)=\text{Re}\delta\omega\left(\mathbf{k}\right)+i\gamma \left(\mathbf{k}\right)$, where with assumption that $\nu_{e}\gg \omega\left(\mathbf{k}\right)$ and $\nu_{e}\gg k_{z}v_{Te}$, 
\begin{eqnarray}
&\displaystyle \text{Re}\, \delta\omega\left(\mathbf{k}\right)=-\frac{\omega^{3}\left(\mathbf{k}\right)k^{2}_{\perp}\rho^{2}_{s}}{\left(1+k^{2}_{\perp}\rho^{2}_{s} \right)k^{2}_{z}v^{2}_{Te}},
\label{48}
\end{eqnarray}
and
\begin{eqnarray}
&\displaystyle \gamma\left(\mathbf{k}\right)=\frac{\nu_{e}\omega^{2}\left(\mathbf{k}\right)
k^{2}_{\perp}\rho^{2}_{s}}{\left(1+k^{2}_{\perp}\rho^{2}_{s} \right)k^{2}_{z}v^{2}_{Te}}\label{49}
\end{eqnarray}
is well known growth rate of the resistive drift instability\cite{Kadomtsev}.
We obtain for $\left(V'_{0}\right)^{-1}<t< t_{s}$ the solution to Eq.(\ref{47}) in the form
\begin{eqnarray}
&\displaystyle \Phi\left(\mathbf{k},t\right)=\Phi_{0}\exp
\left[-i\omega\left(\mathbf{k}\right)\left(t-\frac{\left( 1+T\right)t^{3} }{3\left(T+k^{2}_{\perp}\rho^{2}_{i}\right)t_{s}^{2}b_{i}}
\right)+i\text{Re}\,
\delta\omega\left(\mathbf{k}\right)t\right.\nonumber
\\ &\displaystyle\left.+\left(\gamma\left(\mathbf{k}\right)t
-\frac{t^{2}}{\left(T+k^{2}_{\perp}\rho^{2}_{i}\right)t_{s}^{2}}
\left(1+\frac{T+k^{2}_{\perp}\rho^{2}_{i}}{2b_{i}}\right)\right)\right],
\label{50}
\end{eqnarray}
in which effect of the shear flow is the same as it was obtained in the linear non-modal theory of the 
kinetic drift instability\cite{Mikhailenko-2011} -- shear flow leads to the decrease the 
frequency and the growth rate of the resistive drift instability with time.

In long-time limit, $t\gg t_{s}$, arguments of Bessel function $I_{0}$ in Eq.(\ref{40}) become large, $\hat{k}\left(t\right)\hat{k}\left(t_{1}\right)\rho_{i}^{2}\simeq t^{2}/t^{2}_{s}\gg 1$ and asymptotics (\ref{34})
simplifies Eq.(\ref{40}) for potential $\varphi \left(\mathbf{k}, t\right)$,
\begin{eqnarray}
&\displaystyle \left(1+T \right)\int\limits^{t}_{t_{0}}dt_{1}\frac{d\varphi \left(\mathbf{k}, t\right)}{dt_{t}}=
\int\limits^{t}_{t_{0}}dt_{1}\frac{d}{dt_{t}}\left[\varphi \left(\mathbf{k}, t\right)\frac{t_{s}}
{\sqrt{2\pi t t_{1}}} e^{-\frac{\kappa^{2}_{i}}{2}\left(t-t_{1}\right)^{2}}\right]  \nonumber \\
&\displaystyle +ik_{y}v_{di}\int\limits^{t}_{t_{0}}dt_{1}\varphi \left(\mathbf{k}, t\right)
\frac{t_{s}}{\sqrt{2\pi t t_{1}}}
e^{-\frac{\kappa^{2}_{i}}{2}\left(t-t_{1}\right)^{2}}  \nonumber \\
&\displaystyle +\varphi \left(\mathbf{k}, t_{0}\right)\frac{t_{s}}{t_{0}\sqrt{2\pi}}\left( 
 e^{-\frac{1}{2t^{2}_{s}}\left(t-t_{0}\right)^{2}}- 1
\right),\label{51}
\end{eqnarray} 
where $\kappa^{2}_{i}=t^{-2}_{s}+k^{2}_{z}v^{2}_{Ti}$. At $t\gg t_{s}$,  the right hand side part of 
Eq.(\ref{51}) is proportional to small parameter $t_{s}/t\sim t_{s}/t_{0}$. Applying the methodology 
of the solution of Eq.(\ref{36}) to Eq.(\ref{51}), we obtain the following solution:
\begin{eqnarray}
&\displaystyle
\varphi \left(\mathbf{k}, t\right)=\varphi_{0}\exp \left[\frac{1}{\left(1+T \right)}\frac{t_{s}}{t }
\left(i\frac{k_{y}v_{di}}{2\kappa_{i}}+\sqrt{\frac{2}{\pi}}\frac{1}{k^{2}_{z}
v^{2}_{Ti}t^{2}_{s}} \right)  \right],
\label{52}
\end{eqnarray}
which reproduces qualitatively the same time dependence as of solution (\ref{37}).

\section*{CONCLUSIONS}

In this paper, we develop the non-modal kinetic theory of the hydrodynamic ion temperature gradient 
and resistive drift instabilities in plasma shear flow. In this theory, which is grounded on the shearing modes approach in the kinetic theory of plasma shear flows\cite{Mikhailenko-2011}, the shear flow reveals as the time-dependent effect of the finite Larmor radius in the integral equation for the electrostatic potential. This effect is of principal importance for turbulence evolution in plasma shear flows. It consists in the interaction of ions undergoing 
cyclotron motion with inhomogeneous electric field of sheared modes, which due to 
their distortion by shear flows have time dependent wave number in laboratory frame. 
In Ref.\cite{Mikhailenko-2011} we obtain, that this effect is a source of the enhanced
suppression of the kinetic drift instability by shear flow. In this paper we find, that 
in spite of their hydrodynamic nature, ion temperature gradient and resistive drift 
instabilities in shear flow at times $t>\left(V'_{0}\right)^{-1}$  pass through the 
same linear non-modal kinetic processes of their evolution, as the kinetic drift instability.
These processes reveal in the non-modal decrease with time the 
frequency and the growth rate of the unstable perturbations that display the universality in 
the linear description of the temporal evolution of the electrostatic drift instabilities of shear flows. 

It is important to note, that being qualitatively similar in their linear evolutionary theory, 
these two discussed instabilities have different nonlinear descriptions. The renormalized 
nonmodal nonlinear theory\cite{Mikhailenko-2011}, which accounts for the effect of turbulent 
scattering of ions by the ensemble of the sheared modes with randomly distributed initial 
phases, developed for the kinetic drift instability, is completely applicable to the resistive                                                                                                                             
drift instability and gives completely the same result, 
determined by equation (68) in Ref.\cite{Mikhailenko-2011}, where the frequency and 
the growth rate are determined now by Eqs.(\ref{48}), (\ref{49}). This nonlinear effect,
which is absent in conventional gyrokinetic theory,  consists in the scattering of the ion gyration angle 
by sheared perturbations and reveals in the reduction of the growth rate 
with time as $\gamma\left(V'_{0}t\right)^{-6}$. 

The hydrodynamic ion temperature gradient instability considered here have the modal growth rate 
of the order of its frequency. Therefore,  the methodology of the renormalized nonlinear 
theory, developed in Ref.\cite{Mikhailenko-2011}, which require the smallness of the 
growth rate in comparison with frequency, is not applicable for that instability and 
new  nonlinear theory have to be developed.

\begin{acknowledgments}
This work was funded by National R$\&$D Program through the National
Research Foundation of Korea(NRF) funded by the Ministry of Education, Science and 
Technology (Grant N 2012-034918).
\end{acknowledgments}

\appendix
\section{TRANSFORMATION TO SHEARED COORDINATES FOR NON-STATIONARY SHEAR FLOW}

The procedure of the transformation of the Vlasov-Poisson system to the sheared coordinates, 
developed in Ref.\cite{Mikhailenko-2011}, is generalized easily on the case of the spatially 
homogeneous, but time dependent velocity shear,
\begin{equation}
\mathbf{V}\left(\mathbf{r},
t\right)=\mathbf{V}_{0}\left(x, t\right) =-\frac{c}{B}\frac{\partial
E_{0}\left(x, t\right)}{\partial x}x\mathbf{e}_{y} =\frac{\partial
V_{0}\left(x, t\right)}{\partial x}x\mathbf{e}_{y}=V'_{0}\frac{da\left(t\right)}{dt}x\mathbf{e}_{y}.
\label{A.1}
\end{equation}
where $V'_{0}$  is a parameter with dimension of the velocity shear
and  $a\left(t\right)$ is a function with dimension of time. 
The transition in the Vlasov equation from coordinates $\hat{x}$, $\hat{y}$, $\hat{z}$ and velocity 
$\hat{\mathbf{v}}$ to convected with flow velocity coordinates $\hat{v}_{x}=v_{x}$, $\hat{v}_{y}=v_{y}+V'_{0}(da/dt)x$, $\hat{v}_{z}=v_{z}$, and sheared coordinates, determined by 
$\hat{x}=x$, $\hat{y}=y+V'_{0}a\left(t\right)x$, $\hat{z}=z$ (it is assumed that inhomogeneous electric field, 
and respectively shear flow originate at time $t=0$ and time ordering $\omega_{ci}T\gg 1$ is adapted, 
where $T$ is characteristic time of the flow velocity variations) removes from 
the Vlasov equation spatial inhomogeneity introduced by shear flow. With leading center coordinates 
$X$, $Y$, determined in convective set of reference, 
\begin{equation}
X = x +\frac{v_{\perp}}{\sqrt{\eta\left(t\right)}\omega_{c}}\sin\phi,
\end{equation}
\begin{equation}
Y=y-\frac{v_{\bot}}{\eta\left(t\right)\omega_{c}}\cos
\phi -V'_{0}a\left(t\right)\; \left(X-x\right),
\qquad z_{1}=z-v_{z}t,\label{A.2}
\end{equation}
and velocity space coordinates
\begin{equation}
v_{x}=v_{\bot}\cos \phi,\qquad v_{
y}=\sqrt{\eta} v_{\bot}\sin \phi, \qquad \phi=
\phi_{1}-\omega_{c}\mu\left(t\right), \qquad v_{z}=v_{z},\label{A.3}
\end{equation}
where
\begin{equation}
\eta\left(t\right)=1+\frac{V'_{0}}{\omega_{c}}\frac{da\left(t\right)}{dt},
\qquad
\mu\left(t\right)=\int_{0}^{t}\sqrt{\eta\left(t_{1}\right)}dt_{1},\label{A.4}
\end{equation}
we obtain Vlasov equation in form (\ref{16}) with time dependent coefficient $\eta$. 
The electrostatic potential $\varphi\left(x, y, z, t\right)$ in Eq.(\ref{16}) 
is determined by (\ref{17}), where now
\begin{equation}
\hat{k}^{2}_{\perp}\left(t\right)=\left(k_{x}-V'_{0}a\left(t\right)k_{y}\right)^{2}+\frac{1}{\eta\left(t\right)}k_{y}^{2},\label{A.5}
\end{equation}
and $\tan \theta =k_{y}/\sqrt{\eta\left(t\right)}(k_{x}-V'_{0}a\left(t\right)k_{y})$. 
Therefore, the results for the time 
dependent shear (\ref{A.1}) are reproduced easily from ones obtained for the stationary shear 
by the replacements $V'_{0}t$ on $V'_{0}a\left(t\right)$. For real experimental conditions  
$V'_{0}\ll \omega_{ci}$; so $\eta=1$ and $\mu=1$ and in such case 
Eq.(\ref{16}) still has not explicit time dependence. As a sample of such analysis, we present 
here the extension of the solution Eq.(\ref{50}), obtained with assumption of $\eta=1$, on the time 
dependent velocity shear 
\begin{eqnarray}
&\displaystyle \Phi\left(\mathbf{k},t\right)=\Phi_{0}\exp
\left[-i\omega\left(\mathbf{k}\right)\left(t-\frac{1+T}{\left(T+k^{2}_{\perp}\rho^{2}_{i}\right)
t_{s}^{2}b_{i}}\int^{t}_{0}a^{2}\left(t_{1}\right)dt_{1}\right)+i\text{Re}\,
\delta\omega\left(\mathbf{k}\right)t\right.\nonumber
\\ &\displaystyle\left.+\left(\gamma\left(\mathbf{k}\right)t
-\frac{a^{2}\left(t\right)}{\left(T+k^{2}_{\perp}\rho^{2}_{i}\right)t_{s}^{2}}
\left(1+\frac{T+k^{2}_{\perp}\rho^{2}_{i}}{2b_{i}}\right)\right)\right].\label{A.6}
\end{eqnarray}
The performed analysis displays, that the effects of shear flows becomes appreciable  
when the time-dependent part of the wave number (\ref{A.5}), becomes dominant, i.e. 
when $|V'_{0}a\left(t\right)|\gg 1$. That condition may occur for the steady or for the growing 
with time velocity shear. For the velocity shear (\ref{A.1}) oscillating with time 
as $da\left(t\right)/dt\sim \sin \omega_{0}t$, we have $|V'_{0}a\left(t\right)|\lesssim V'_{0}/\omega_{0}$
and condition $|V'_{0}a\left(t\right)|\gg 1$ may be attained only for low frequency 
oscillation of the velocity shear, when the condition $|V'_{0}|\gg \omega_{0}$ is met; 
effect of shear flow is negligible for $\omega_{0}>|V'_{0}|$.

\end{document}